\documentclass[11pt]{article} 
\usepackage{latexsym}

\def\hybrid{\topmargin -25pt    \oddsidemargin 0in 
	\headheight 0pt \headsep 0pt
	\textwidth 6.5in       
	\textheight 9.5in       
	\marginparwidth .875in
	\parskip 5pt plus 1pt   \jot = 1.5ex}
\hybrid
\def\cQ{{\cal Q}}
\def\cG{{\cal G}}
\def\cL{{\cal L}}

\def\cM{{\cal M}}
\def\cH{{\cal H}}
\def\ket#1{|{#1}\rangle}
\def\noi{\noindent}

\def\baselinestretch{1.2}

\catcode`\@=11

\def\marginnote#1{}
\def\draftlabel#1{{\@bsphack\if@filesw {\let\thepage\relax
   \xdef\@gtempa{\write\@auxout{\string
      \newlabel{#1}{{\@currentlabel}{\thepage}}}}}\@gtempa
   \if@nobreak \ifvmode\nobreak\fi\fi\fi\@esphack}
	\gdef\@eqnlabel{#1}}
\def\@eqnlabel{}
\def\@vacuum{}
\def\draftmarginnote#1{\marginpar{\raggedright\scriptsize\tt#1}}

\def\draft{\oddsidemargin -.2truein
	\def\@oddfoot{\sl preliminary draft \hfil
	\rm\thepage\hfil\sl\today\quad\militarytime}
	\let\@evenfoot\@oddfoot \overfullrule 3pt
	\let\label=\draftlabel
	\let\marginnote=\draftmarginnote
   \def\@eqnnum{(\theequation)\rlap{\kern\marginparsep\tt\@eqnlabel}%
\global\let\@eqnlabel\@vacuum}  }

\def\preprint{\twocolumn\sloppy\flushbottom\parindent 2em
	\leftmargini 2em\leftmarginv .5em\leftmarginvi .5em
	\oddsidemargin -.5in    \evensidemargin -.5in
	\columnsep .4in \footheight 0pt
	\textwidth 10.in        \topmargin  -.4in
	\headheight 12pt \topskip .4in
	\textheight 6.9in \footskip 0pt
	\def\@oddhead{\thepage\hfil\addtocounter{page}{1}\thepage}
	\let\@evenhead\@oddhead \def\@oddfoot{} \def\@evenfoot{} }

\def\numberbysection{\@addtoreset{equation}{section}
	\def\theequation{\thesection.\arabic{equation}}}

\def\underline#1{\relax\ifmmode\@@underline#1\else
	$\@@underline{\hbox{#1}}$\relax\fi}

\def\titlepage{\@restonecolfalse\if@twocolumn\@restonecoltrue
\onecolumn
     \else \newpage \fi \thispagestyle{empty}\c@page\z@
	\def\thefootnote{\fnsymbol{footnote}} }

\def\endtitlepage{\if@restonecol\twocolumn \else \newpage \fi
	\def\thefootnote{\arabic{footnote}}
	\setcounter{footnote}{0}}  

\catcode`@=12
\relax

\def\figcap{\section*{Figure Captions\markboth
	{FIGURECAPTIONS}{FIGURECAPTIONS}}\list
	{Figure \arabic{enumi}:\hfill}{\settowidth\labelwidth{Figure
999:}
	\leftmargin\labelwidth
	\advance\leftmargin\labelsep\usecounter{enumi}}}
 \relax
\def\tablecap{\section*{Table Captions\markboth
	{TABLECAPTIONS}{TABLECAPTIONS}}\list
	{Table \arabic{enumi}:\hfill}{\settowidth\labelwidth{Table
999:}
	\leftmargin\labelwidth
	\advance\leftmargin\labelsep\usecounter{enumi}}}
 \relax
\def\reflist{\section*{References\markboth
	{REFLIST}{REFLIST}}\list
	{[\arabic{enumi}]\hfill}{\settowidth\labelwidth{[999]}
	\leftmargin\labelwidth
	\advance\leftmargin\labelsep\usecounter{enumi}}}
 \relax
\makeatletter
\newcounter{pubctr}
\def\publist{\@ifnextchar[{\@publist}{\@@publist}}
\def\@publist[#1]{\list
	{[\arabic{pubctr}]\hfill}{\settowidth\labelwidth{[999]}
	\leftmargin\labelwidth
	\advance\leftmargin\labelsep
	\@nmbrlisttrue\def\@listctr{pubctr}
	\setcounter{pubctr}{#1}\addtocounter{pubctr}{-1}}}
\def\@@publist{\list
	{[\arabic{pubctr}]\hfill}{\settowidth\labelwidth{[999]}
	\leftmargin\labelwidth
	\advance\leftmargin\labelsep
	\@nmbrlisttrue\def\@listctr{pubctr}}}
 \relax
\makeatother
\newskip\humongous \humongous=0pt plus 1000pt minus 1000pt

\newif\ifdtup

\font\Scbig=cmss10 scaled\magstep1
\font\Scscr=cmss8 scaled\magstep1
\font\Scscrscr=cmss8
\newfam\Scfam
\textfont\Scfam=\Scbig
\scriptfont\Scfam=\Scscr
\scriptscriptfont\Scfam=\Scscrscr
\def\Sc{\fam\Scfam}

\relax
\def\lvm{\leavevmode\hbox to\parindent{\hfill}}

\def\BE{\begin{equation}}
\def\EE{\end{equation}}
\def\BA{\begin{eqnarray}}
\def\EA{\end{eqnarray}}

\def\D{\Delta}

\def\tt{\bar\tau}

\def\lvm{\leavevmode\hbox to\parindent{\hfill}}
\def\bar{\overline}

\def\L{\left}
\def\R{\right}

\def\BE{\begin{equation}}
\def\EE{\end{equation} \vskip 0.30\baselineskip}
\def\BA{\begin{array}}
\def\EA{\end{array}}

\def\noi{\noindent}

\def\frac#1#2{{\textstyle{{#1}\over{#2}}}}

\def\Kr#1{\delta_{{#1},0}}
\def\ket#1{|{#1}\rangle}

\def\cG{{\cal G}}
\def\cH{{\cal H}}

\def\cL{{\cal L}}

\def\cQ{{\cal Q}}

\def\open#1{\mbox{{\bf{#1}}}}

\def\oZ{{\open Z}}

\def\ctop{{\Sc c}}
\def\htop{{\Sc h}}

\def\svec{singular vector}

\def\Qz{\cQ_0}
\def\Gz{\cG_0}
\def\Qn{$\Qz$}
\def\Gn{$\Gz$}
\def\kc{{\ket{\chi}}}
\def\kcc#1#2#3{{\kc_{#1}^{({#2}){#3}}}}

\newif\ifold \oldtrue \def\new{\oldfalse}
\let\ssection=\section
\def\section{\setcounter{equation}{0}\ssection}

\begin{document}
\renewcommand{\theequation}{\thesection.\arabic{equation}}
\newcommand{\beq}{\begin{equation}}
\newcommand{\eeq}[1]{\label{#1}\end{equation}}
\newcommand{\bea}{\begin{eqnarray}}
\newcommand{\eea}{\end{eqnarray}}
\newcommand{\eer}[1]{\label{#1}\end{eqnarray}}
\begin{titlepage}
\begin{center}

\hfill IFF-FM-2008/02\\
\hfill Nikhef-2008-013\\
\vskip 1in

{\large \bf 
No isomorphism between the affine $\hat sl(2)$ algebra and the $N=2$ superconformal
algebras}                        
\vskip .7in

 {\bf Beatriz Gato-Rivera}\footnote{Also known as B. Gato} \\
\vskip .6in

{\em Instituto de F\'\i sica Fundamental, CSIC \\
Serrano 123, Madrid 28006, Spain}\\

{\em NIKHEF-H, Kruislaan 409, NL-1098 SJ Amsterdam, The Netherlands}\\

\vskip 1.3in

\end{center}

\begin{center} {\bf ABSTRACT } \end{center}
\begin{quotation}
Since 1999 it became obvious that the would be `isomorphism' between 
the affine $\hat sl(2)$ algebra and the N=2 superconformal algebras, proposed
by some authors, simply does not work. However, this issue was never
properly discussed in the literature and, as a result, some confusion still
remains. In this article we finally settle down, clearly and unambiguously,
the true facts: there is no isomorphism between the affine $\hat sl(2)$ 
algebra and the N=2 superconformal algebras. 
\end{quotation}
\vskip 1in

September 2008
\end{titlepage}

\def\baselinestretch{1.2}
\baselineskip 16 pt
\section{Introduction}\lvm

Subsingular vectors of the N=2 superconformal algebras were discovered, and 
examples given, in 1996 \cite{BJI6} \cite{BJI5}. Shortly afterwards 
Semikhatov and Tipunin claimed to have obtained a complete classification 
of the N=2 subsingular vectors in the paper `The Structure of Verma Modules 
over the N=2 Superconformal algebra' \cite{SeTi3}. Surprisingly, the only 
explicit examples of N=2 subsingular vectors known at that time did not fit
into their classification. All the results presented in that paper, including
the classification of subsingular vectors, were based on the assumption
that there exists an isomorphism between the affine $\hat sl(2)$ algebras and the 
$N=2$ superconformal algebras, proposed by the authors in earlier work
\cite{SeTi1}\cite{SeTi2} without proofs. Using this `isomorphism' the authors: i) deduced 
that there were only two different types of submodules in N=2 Verma modules, 
and ii) claimed that they had constructed `non-conventional' singular 
vectors with the property of generating the two types of submodules maximally, i.e.
with no subsingular vectors left outside. The classification of the N=2 
subsingular vectors then followed applying these two results. 
A couple of years later, in 1999, after some more papers by the same authors 
had appeared making use of the `isomorphism', we proved, in a note sent to 
the archives \cite{B2}, that both results were incorrect: there are four different 
types of submodules in N=2 Verma modules and the `non-conventional'
vectors do not generate the submodules maximally (we used one explicit
example to see this). However, we did not emphasize enough the fact that
our results provided a strong indication that the `would be  isomorphism'
was incorrect, we just made a small comment about it.  Although the authors 
did not make use of the `isomorphism' again (as far as we know!), the lack of 
a clear discussion about this issue brought consequences: the 
`isomorphism' remained, and still remains, as a belief by some colleagues
who have not worked out the details, as we have had the chance to 
verify in several occasions, either in writing form or verbally. It is our
intention now to clarify this issue, finally, and to settle down the true
facts: there is no isomorphism between the affine $\hat sl(2)$ algebra 
and the N=2 superconformal algebras. 

Our strategy will be to prove that
the predictions of this  `would be  isomorphism' regarding the Topological 
N=2 superconformal algebra, which is the N=2 algebra considered by 
Semikhatov and Tipunin, simply do not work. We do not find  necessary 
to make a detailed analysis of the `isomorphism' itself trying to understand
why it does not work. The authors never provided a proof that their construction
provided an isomorphism!\footnote{This was very much their style. For example,
in ref. \cite{SeTi3} they wrote eleven theorems, seven propositions, four
lemmas and zero proofs (not even low level explicit examples).}
Otherwise we would have felt somehow compelled to find the mistakes. 
In what follows, in section 2 we will first introduce the Topological N=2 algebra,
together with some important results regarding its representation theory. In 
section 3 we will describe the major claims made by Semikhatov and Tipunin
in several papers (the first ones being \cite{SeTi1} \cite{SeTi2} \cite{SeTi3}), 
regarding the different types and properties of the submodules of the 
Topological N=2 algebra, as deduced directly from the `isomorphism' 
between this algebra and the affine $\hat sl(2)$ algebra. Then in section 4 
we will prove that these claims are incorrect.

\section{The Topological N=2 superconformal algebra}\lvm

The Topological N=2 superconformal algebra was deduced in 1990
as the symmetry algebra of two-dimensional topological conformal field 
theory (TCFT) \cite{DVV}. It was the
last N=2 superconformal algebra to be discovered and in fact can be obtained
from the Neveu-Schwarz N=2 algebra by modifying 
the stress-energy tensor by adding the derivative of the U(1) current, a
procedure known as {\it topological twist} \cite{EY}\cite{W-top}. It reads
\BE\new\BA{lclclcl}
\L[\cL_m,\cL_n\R]&=&(m-n)\cL_{m+n}\,,&\qquad&[\cH_m,\cH_n]&=
&{\ctop\over3}m\Kr{m+n}\,,\\
\L[\cL_m,\cG_n\R]&=&(m-n)\cG_{m+n}\,,&\qquad&[\cH_m,\cG_n]&=&\cG_{m+n}\,,
\\
\L[\cL_m,\cQ_n\R]&=&-n\cQ_{m+n}\,,&\qquad&[\cH_m,\cQ_n]&=&-\cQ_{m+n}\,,\\
\L[\cL_m,\cH_n\R]&=&\multicolumn{5}{l}{-n\cH_{m+n}+{\ctop\over6}(m^2+m)
\Kr{m+n}\,,}\\
\L\{\cG_m,\cQ_n\R\}&=&\multicolumn{5}{l}{2\cL_{m+n}-2n\cH_{m+n}+
{\ctop\over3}(m^2+m)\Kr{m+n}\,,}\EA\qquad m,~n\in\oZ\,.\label{topalgebra}
\EE
\noi
where $\cL_m$ and $\cH_m$ are the bosonic generators corresponding to the
stress-energy tensor (Virasoro generators) and the U(1) current,
respectively, and $\cG_m$ and $\cQ_m$ are the spin-2 and spin-1 fermionic 
generators, the latter being the modes of the BRST-current.
The eigenvalues of the bosonic zero modes $(\cL_0,\,\cH_0)$ correspond to
the conformal weight and the U(1) charge of the states. 
In a Verma module these eigenvalues split
conveniently as $(\D+l,\,\htop+q)$ for secondary states, where $l$
and $q$ are the {\it level} and the {\it relative charge} of the state and
$(\D,\,\htop)$ are the conformal weight and the charge of
the primary state on which the secondary is built.
The `topological' anomaly $\ctop$ is the conformal anomaly 
corresponding to the Neveu-Schwarz N=2 algebra.

Due to the existence of the fermionic zero modes \Gn\ and \Qn\  
this algebra has two sectors: the $\cG$-sector (states annihilated 
by \Gn ) and the $\cQ$-sector (BRST-invariant states annihilated by 
\Qn ), in analogy with the $(+)$-sector and the $(-)$-sector of the 
Ramond N=2 algebra, due to the fermionic zero modes $G_0^+$ 
and $G_0^-$. (As a matter of fact, these two N=2 algebras are
exactly isomorphic, as was proven in \cite{DB4}.) However,
the two sectors do not provide the complete description 
since there are also states which do not belong to any of the sectors
\cite{BJI6}\cite{DB2}\cite{DB4}. That is, not all Verma modules and
submodules decompose into the two sectors, but there are also 
indecomposable states, in particular indecomposable \svec s.
To see this one only needs to inspect the anticommutator of the
fermionic zero modes $\{\Gz , \Qz \}=2\cL_0$ acting on a given state $\kc$. 
If the conformal weight of $\kc$ is different from zero; i.e.
$\cL_0 \kc = (\D+l) \kc \neq 0$, then $\kc$ can be decomposed into a
state $\kc^G$ annihilated by \Gn , but not by \Qn , that we refer as
\Gn-closed and a state $\kc^Q$ annihilated by \Qn , 
but not by \Gn , that we refer as \Qn-closed:
\BE \kc = {1\over 2\D} \Qz \Gz \kc + {1\over 2\D} \Gz \Qz \kc = 
\kc^Q + \kc^G    \,. \EE   

If the conformal weight of $\kc$ is zero, however, one only obtains
$(\Gz \Qz + \Qz \Gz)\kc=0$, which is satisfied in four different ways: 
i) The state is \Gn-closed, $\kc=\kc^G$, and $\Gz \Qz \kc^G =0$,
ii) The state is \Qn-closed, $\kc=\kc^Q$, and $\Qz \Gz \kc^Q =0$,
iii) The state is chiral, $\kc=\kc^{G,Q}$, annihilated by both \Gn\ and \Qn ,
and iv) The state is indecomposable `no-label', $\kc=\kc$, not annihilated
by any of the fermionic zero modes. 
   
In what follows we will use the standard definition of 
highest weight vectors and \svec s for conformal 
algebras, i.e. they are the states with {\it lowest} conformal weight
({\it lowest} energy) in the Verma modules and in the null submodules, 
respectively, and therefore are annihilated by all the positive
modes of the generators of the algebra (the lowering operators); i.e. 
${\ } \cL_{n \geq 1} \kc =  \cH_{n \geq 1} \kc =  {\cG}_{n \geq 1} \kc
=  {\cQ}_{n \geq 1} \kc = 0 {\ }$. Hence these annihilation conditions will be 
referred to as the conventional, standard highest weight (h.w.) conditions. 
Singular vectors that are not generated by acting with the algebra on 
other \svec s are called {\it primitive}, otherwise they are called
{\it secondary} \svec s.

Subsingular vectors are also null but they do not satisfy the h.w.
conditions, becoming singular, that is annihilated by all the 
positive generators, in the quotient of the Verma module by a
submodule, however. As a consequence they are 
located {\it outside} that particular submodule (otherwise they would
disappear after taking the quotient), although descending to it 
necessarily by the action of the lowering operators (so that they
descend to `nothing' once the submodule is set to zero). 
This implies that
{\it the \svec s cannot reach the subsingular vectors } going upwards by 
the action of the negative, rising operators, whereas {\it the subsingular
vectors can reach the \svec s } going downwards by the action of the
positive, lowering operators.   

Subsingular vectors for the N=2 algebras were discovered in 1996 in
ref. \cite{BJI5} and the first examples for the case of the Topological N=2
algebra were published in January 1997 in ref. \cite{BJI6}, together with
the classification of all possible types of singular vectors taking into
account the relative U(1) charge and the annihilation conditions with 
respect to the fermionic zero modes \Gn\ and \Qn\ .
This classification resulted in: 4 different types of \svec s
for chiral Verma modules built on chiral highest weight vectors
$\ket{0,\htop}^{G,Q}$, 10 different types of \svec s for 
generic (standard) Verma modules built on 
\Gn-closed h.w. vectors $\ket{\D,\htop}^G$, another 10 types for
generic Verma modules built on \Qn-closed h.w. vectors 
$\ket{\D,\htop}^Q$,  and 9 different types of \svec s for 
`no-label'  Verma modules built on indecomposable
h.w. vectors $\ket{0,\htop}$. In generic Verma modules one can find
\Gn-closed, \Qn-closed, chiral and  indecomposable \svec s. 
In chiral and no-label Verma modules, however, only \Gn-closed and 
\Qn-closed \svec s can exist, with the exception of the chiral \svec s
at level zero in no-label Verma modules. For the case of the generic 
Verma modules built on \Gn-closed h.w. vectors $\ket{\D,\htop}^G$,
which were the only generic Verma modules considered by 
Semikhatov and Tipunin (they ignored the ones built  on \Qn-closed 
h.w. vectors as well as the no-label Verma modules built on
indecomposable h.w. vectors), the possible types of \svec s 
one can find are given by the following 
table  \cite{BJI6}  \cite{DB2} :

\BE
\begin{tabular}{r|l l l l}
{\ }& $q=-2$ & $q=-1$ & $q=0$ & $q=1$\\
\hline\\
\Gn-closed & $-$ & $\kc_l^{(-1)G}$ & $\kc_l^{(0)G}$ & $\kc_l^{(1)G}$\\
\Qn-closed & $\kc_l^{(-2)Q}$ & $\kc_l^{(-1)Q}$ & $\kc_l^{(0)Q}$ & $-$ \\
chiral & $-$ & $\kc_l^{(-1)G,Q}$ & 
$\kc_l^{(0)G,Q}$ & $-$ \\
indecomposable & $-$ & $\kc_l^{(-1)}$ &
$\kc_l^{(0)}$ & $-$\\
\end{tabular}
\label{tabl2}
\EE
                             
In ref. \cite{BJI6} all \svec s (i.e. 4 + 20 + 9)
were written down explicitly at level 1. This classification of \svec s
was proven to be rigorous later in ref. \cite{DB2}, using the results for the
maximal dimensions of the corresponding spaces of singular vectors 
(1, 2 or 3 depending on the type of singular vector). 
Regarding subsingular vectors, in ref. \cite{BJI6} 
all the subsingular vectors in generic Verma modules  
that become singular in the chiral Verma modules were written down 
at levels 2 and 3. To understand this
one has to take into account that chiral Verma modules are nothing but
the quotient of generic Verma modules with zero conformal weight, $\D=0$,
by the submodules generated by the level-zero \svec s (which are present 
in all generic Verma modules with $\D=0$). 

\section{The Claims}\lvm

Three months after the paper \cite{BJI6} was published in the archives
(January 97), the paper  \cite{SeTi3} appeared 
also in the archives. As was mentioned before, in \cite{SeTi3} as
well as in earlier work the authors considered only the Topological N=2 
algebra (among the four existing N=2 algebras). All the 
analysis and results presented by the authors were based on the 
claim that there exists an isomorphism between the affine $\hat sl(2)$ 
algebra and the N=2 superconformal algebra at hand, giving rise to the
following two major assumptions that were described as proven facts:
 
i) In the N=2 Verma modules there are only two types of submodules. In
particular, in the generic Verma modules built on \Gn-closed h.w. vectors
(called `massive' Verma modules by the authors) one can find the two 
types of submodules, denoted as `massive' (large) and `topological' (small).

Let us notice already that, although ref. \cite{BJI6} appeared in the 
bibliography given by the authors, the classification of Verma modules 
(generic, no-label and chiral), with their possible existing types 
of \svec s, was overlooked. In particular the authors ignored the
indecomposable \svec s in generic (`massive') Verma modules
(see table (1.3)),  which clearly generate a different type of submodule 
with no counterpart in the affine $\hat sl(2)$ algebra, as we will show. 
In other words, the very existence of the indecomposable \svec s of the 
Topological N=2 algebra, written down explicitely at level 1  in ref.  \cite{BJI6}, 
lacking of a counterpart in the affine $\hat sl(2)$ algebra, is already sufficient
to disprove any possible isomorphism between these two algebras.

ii) These two types of submodules are maximally generated (i.e. without 
letting any null states outside, like subsingular vectors) by some 
`non-conventional \svec s', constructed by the authors in 
refs. \cite{SeTi1} \cite{SeTi2}, which satisfy `twisted' h.w. conditions and 
coincide with the conventional \svec s only in the case of `zero twist'.
In more intuitive terms one can think of the `non-conventional' \svec s
simply as certain null states which, unlike the conventional \svec s,
are not located at the bottom of the submodules, that is, they are not the 
null states with lowest conformal weight, except for the case of `zero twist'.
(In our opinion, the authors use an unnecessary complicated notation:
null states that generate bigger submodules than the h.w.  \svec s are
nothing but subsingular vectors).
   
Based on these assumptions the authors presented a `complete' classification
of subsingular vectors for the Topological N=2 algebra (without giving explicit 
examples) where, surprisingly, the subsingular vectors given in ref. \cite{BJI6}, 
which were the only explicit examples written down so far, did not fit. If one takes
into account that subsingular vectors do not exist for the affine $\hat sl(2)$ algebra, 
one might wonder whether the discovery of subsingular vectors for the N=2
superconformal algebras should have been already a strong indication against
the existence of an isomorphism between these algebras and  the affine 
$\hat sl(2)$ algebra. The strategy of the authors then was to claim that the
isomorphism mapped the Verma modules and submodules of the two algebras 
between each other, in such a way that the singular vectors of the  
$\hat sl(2)$ algebra, which generate the submodules maximally in the absence 
of subsingular vectors, would correspond to the `non-conventional \svec s' 
of the Topological N=2 algebra, the N=2 subsingular vectors corresponding to
some ordinary null states of $\hat sl(2)$ (or to singular vectors in the special
case in which the N=2 subsingular vectors and the `non-conventional \svec s'
coincide). An important observation is that the standard Verma modules of
the $\hat sl(2)$ algebra were supposed to be isomorphic to the chiral Verma
modules of the Topological N=2 algebra, built on chiral h.w. vectors
$\ket{0,\htop}^{G,Q}$, that the authors called `topological Verma modules'. In order
to construct the isomorphic counterparts of the generic N=2 Verma modules,
the authors defined the `relaxed Verma modules' built on non-standard h.w.
vectors of $\hat sl(2)$.  

\section{The Facts}

In what follows, in subsections 4.1 and 4.2, we will show that:

i) In generic (`massive') Verma modules one can find four
different types of submodules with respect to their size and shape
at the bottom/top.\footnote{We draw the Verma modules from the 
bottom upwards, Semikhatov and Tipunin draw them downwards.} 
Two of them fit, in principle, into the description of `massive' 
and `topological' submodules given by Semikhatov and Tipunin. 
The other two types do not fit into that description. 

ii) The N=2 subsingular vectors written down in ref. \cite{BJI6} do not
fit into the classification presented by the authors in ref. \cite{SeTi3}, 
providing in fact a proof that the `non-conventional \svec s'  do not
generate maximal submodules since one can find subsingular vectors
outside which are pulled inside the submodule by the action of the
positive lowering operators.

\vskip 0.5in
  
\subsection{Different types of submodules}\lvm

The determinant formulae for the Topological N=2 algebra were presented 
in \cite{BJI7} for the chiral (`topological') Verma modules, and in \cite{DB4} for the 
generic (`massive') and `no-label' Verma modules, together with a very detailed
analysis of the \svec s corresponding to the roots of the determinants. In addition 
it was proved -- both theoretically and with explicit examples -- that in generic 
Verma modules one can find four different types of submodules just by taking into
account the size and the shape at the bottom of the submodules. Now we will
review these results and argue that two of these types of submodules do not 
fit into the `massive' and `topological' submodules of Semikhatov and Tipunin,
which according to the `isomorphism' should be the only existing types 
of submodules of the Topological N=2 algebra.
The argument goes as follows. The determinant formula for all 
the generic Verma modules  -- either with two h.w. vectors  
$\ket{\D,\htop}^G$ and $\ket{\D,\htop-1}^Q$ ($\D \neq 0$) or with 
only one h.w. vector $\ket{0,\htop}^G$ or $\ket{0,\htop-1}^Q$ --
reads\footnote{The Verma modules built on \Gn-closed h.w. vectors 
and the ones built on \Qn-closed h.w. vectors are not the same
for zero conformal weight $\D=0$ because in this case there is only one
h.w. vector at the bottom of the Verma module together with one \svec .}
 
\BE
det(\cM_l^T)= 
\prod_{2\leq rs \leq 2l}(f_{r,s})^{2 P(l-{rs\over2})}
\prod_{0 \leq k\leq l}(g_k^+)^{2 P_k(l-k)} 
\prod_{0 \leq k\leq l}(g_k^-)^{2 P_k(l-k)}  {\ } ,
\label{det1}
\EE
\noi
where  
\BE
f_{r,s}(\D,\htop,t) = -2t\D + t\htop - \htop^2 - {1 \over 4} t^2
       + {1 \over 4} (s-tr)^2  \,, {\ \ \ }r\in\oZ^+,{\ }s\in2\oZ^+
       \label{frs} \EE
\noi
and 
\BE
g_k^{\pm}(\D,\htop,t) = 2 \D \mp 2k \htop - tk(k \mp 1) \,, {\ \ \ }
	 0 \leq k \in \oZ \,, \label{gk} \EE
\noi
defining the parameter $t={(3-\ctop) / 3}$.
For $\ctop \neq 3 {\ \ } (t \neq 0) {\ }$ one can factorize $f_{r,s}$ as
\BE
f_{r,s}(\D,\htop,t\neq 0) = -2t (\D-\D_{r,s}) \,, \qquad
\D_{r,s} = - {1\over 2t} (\htop-\htop_{r,s}^{(0)}) (\htop-\hat\htop_{r,s})\,,
\label{Drs} \EE
\noi
with
\BE \htop_{r,s}^{(0)} = {t \over 2}(1+r)-{s \over 2} \ , \ \ \ \ \ 
r \in {\bf Z}^+ , \ \ s \in 2{\bf Z}^+ \,, \label{h0rs} \EE
\BE \hat\htop_{r,s} = {t \over 2}(1-r)+{s \over 2}  \ , \ \ \ \ \ 
r \in {\bf Z}^+ , \ \ s \in 2{\bf Z}^+ \,. \label{hhrs} \EE
\noi
For all values of $\ctop$ one can factorize $g_k^+$ and $g_k^-$ as
\BE g_k^{\pm}(\D,\htop,t) = 2 (\D-\D_k^{\pm}) \,, \qquad
\D_k^{\pm} = \pm k \, (\htop-\htop_{k}^{\pm}) \,,  \label{Dk} \EE
\noi
with 
\BE \htop_k^{\pm} =  {t \over 2}\,(1 \mp k ) \,, \ \ \ \ \ k\in {\bf Z}^+ 
\label{hpml} \EE

\noi
The partition functions are defined by
\BE
\sum_N P_k(N)x^N={1\over 1+x^k}{\ }\sum_n P(n)x^n=
{1\over 1+x^k}{\ }\prod_{0<r\in {\bf Z},{\ }0<m\in {\bf Z}}{(1+x^r)^2
 \over (1-x^m)^2} .
\label{part}\EE
The fact that $2 P(0) = 2 P_k(0) = 2$ indicates that the \svec s 
come two by two at the same level, in the same Verma module. 
Generically one is in the $\cG$-sector,
annihilated by (at least) \Gn , while the other is in the $\cQ$-sector, 
annihilated by (at least) \Qn .  Now comes an observation for the readers 
who are more acquainted with the Neveu-Schwarz N=2 algebra.
The roots of the quadratic vanishing surface $f_{r,s}(\D,\htop,t)=0$
and of the vanishing planes $g_k^{\pm}(\D,\htop,t)=0$
are related to the corresponding roots of the determinant
formula for the Neveu-Schwarz N=2 algebra 
\cite{BFK}\cite{Nam}\cite{KaMa3}\cite{Yu} via the topological twists. 
These transform the standard h.w. vectors of the 
Neveu-Schwarz N=2 algebra into \Gn-closed h.w. vectors of the
Topological N=2 algebra. As a consequence, under the topological
twists, the Neveu-Schwarz \svec s are transformed into the \svec s 
of the $\cG$-sector of the Topological algebra (see refs.
\cite{BJI6}\cite{BJI7} for a detailed account of the twisting and 
untwisting of primary states and singular vectors).

It is easy to check, by counting of states, that the partitions
$2 P(l-{rs\over2})$, exponents of $f_{r,s}$ in the determinant formula, 
correspond to submodules of generic type, of the same size as
the Verma module itself, so to speak, whereas the partitions 
$2 P_k(l-k)$, exponents of $g_k^{\pm}$ in the determinant formula,
correspond to smaller submodules.
Furthermore, as pointed out before, taking into account also 
the shape at the bottom one can distinguish four types of submodules. 
Two of these types correspond to the quadratic vanishing surfaces
$f_{r,s}(\D,\htop,t)=0$, a third type corresponds to the vanishing planes 
$g_{k}^{\pm}(\D, \htop, t)=0$, and the fourth type corresponds to the
`no-label' submodules, built on indecomposable \svec s, that one finds
in certain intersections of $f_{r,s}(\D,\htop,t)=0$ and 
$g_{k}^{\pm}(\D, \htop, t)=0$, as we will explain.

The two types of submodules that correspond to the quadratic vanishing 
surfaces $f_{r,s}(\D,\htop,t)=0$ have therefore the same size, but they 
differ on the shape at 
the bottom, where they both have (in the case $\D \neq 0$)
two {\it uncharged} \svec s at level $l={rs\over2}$: $\kc_l^{(0)G}$ in the
$\cG$-sector and $\kc_l^{(0)Q}$ in the $\cQ$-sector\footnote{
An important technical remark is that if one chooses as h.w. vector of the
Verma module only the \Gn-closed one $\ket{\D,\htop}^G$, as Semikhatov 
and Tipunin do, regarding the \Qn-closed h.w. vector $\ket{\D,\htop-1}^Q$
simply as a descendant state, then the uncharged \svec s $\kc_l^{(0)Q}$ in the 
$\cQ$-sector are necessarily described as negatively charged \svec s 
$\kc_l^{(-1)Q}$ built on the  \Gn-closed h.w. vector $\ket{\D,\htop}^G$. For the 
case $\D=0$ there is only one h.w. vector in the Verma module and therefore
only one of the \svec s can be described as `uncharged' while the other
must necessarily be described as charged with respect to the unique h.w. 
vector.}.
As shown in Figure I, on the left and in the center, in most cases 
the bottom of the submodule consists of two \svec s 
connected by one or two horizontal arrows corresponding to the 
action of  \Qn\ and/or \Gn . There is only one arrow if one of the 
\svec s is chiral, i.e. of type $\kc_l^{(0)G,Q}$ instead, what happens
generically for $\D = -l$. These submodules fit, in principle,
into the description of `massive' submodules given by Semikhatov and 
Tipunin. Namely, `massive' submodules are supposed to correspond to 
the uncharged roots $f_{r,s}(\D,\htop,t)=0$, they have the same size as 
the generic (`massive') Verma module and they have two states
at the bottom connected through \Qn\ and/or \Gn , one of these states 
being the \Gn-closed uncharged \svec\ $\kc_l^{(0)G}$ (they do not
mention the possibility that this \svec\ may be chiral for $\D = -l$, though).

It also happens, however, for $\D = -l, \, t = - {s \over n}, \, n=1,..,r$,
that the two \svec s at the bottom of the submodule are chiral both, 
and therefore disconnected from each other, as shown in Fig. I, on
the right. Consequently these `chiral-chiral' submodules, 
of the same size as the `massive' submodules and corresponding
also to the uncharged roots $f_{r,s}(\D,\htop,t)=0$, contain two 
disconnected pieces at the bottom and as a result do not fit into 
the description of `massive' submodules. Nor do they fit into the 
description of two `topological' (smaller) submodules together since 
these correspond to the charged roots $g_{k}^{\pm}(\D, \htop, t)=0$ with
also two \svec s at the bottom of the submodules connected by the action 
of  \Qn\ and/or \Gn : the charged \svec s $\kc_l^{(1)G}$ or $\kc_l^{(-1)G}$ 
in the $\cG$-sector plus their companions in the $\cQ$-sector, 
as we will see.

Let us stress that the existence of `chiral-chiral' submodules was obvious 
since January 1997 when the whole set of \svec s of the Topological algebra
at level 1 was written down in ref. \cite{BJI6}. For example, the chiral 
\svec s $\kc_1^{(q)G,Q}$ at level 1 built on \Gn-closed h.w. vectors
$\ket{\D,\htop}^G$ (which are the only h.w. vectors considered by 
Semikhatov and Tipunin) were shown to be:
\BE
\kc_1^{(0)G,Q}=(-2\cL_{-1} + \cG_{-1}\cQ_0 )\ket{-1,-1}^G ,\EE
\BE
\kc_1^{(-1)G,Q}=(\cL_{-1}\cQ_0+\cH_{-1}\cQ_0 +
\cQ_{-1}) \ket{-1,{6-\ctop\over3}}^G . \EE
For $\ctop=9$ ($t=-2$) these two chiral \svec s are together in the same 
generic (`massive') Verma module built on the h.w. vector $\ket{-1,-1}^G$.
Hence this example already proves the existence of `chiral-chiral' submodules 
at level 1. (Observe what we indicated in footnote 3: the uncharged \svec s 
$\kc_l^{(0)Q}$ in the $\cQ$-sector are necessarily described as  negatively 
charged  \svec s  $\kc_l^{(-1)Q}$ when built on \Gn-closed h.w. vectors 
$\ket{\D,\htop}^G$. In this example, the uncharged \svec s in the $\cG$-sector 
and in the $\cQ$-sector turn out to be chiral, i.e; annihilated by both  
\Gn\ and \Qn .)

\vskip .3in
\baselineskip 12pt

\vskip .5in
\def\sk {{\hskip 1 cm}}
\def\bk {{\hskip 0.2 cm}}
\def\bbk{{\hskip 0.2 cm}}
\def\rk {{\hskip 3 cm}}
\def\com{{\hskip 0.2 cm},}
\def\pkt{{\hskip 0.2 cm}.}

\def\cQ{{\cal Q}}
\def\cG{{\cal G}}

\def\abs{\,|\,}                                 
\def\tilt {\tilde{t}}             
\def\tils {\tilde{s}}             

\newcommand{\fig}[1]{{\sc Fig.}\,{\sf #1}}        
\newcommand{\figs}[1]{{\sc Figs.}\,{\sf #1}}      
\newcommand{\ch}[1]{\makebox(0,0)[b]{\scriptsize$#1$}}  
\newcommand{\cl}{tt}              
\newcommand{\case}[1]{\makebox(0,0)[b]{\footnotesize #1}}                

\newcounter{pics}
\newcommand{\bpic}[4]{\begin{center}\begin{picture}(#1,#2)(#3,#4)
\refstepcounter{pics}}
\renewcommand{\thepics}{{\sf\Roman{pics}}}
\newcommand{\epic}[1]{\end{picture}\\
\end{center}{ \footnotesize {\sc Fig.} \thepics \bk #1}}
\newcommand{\epicspl}{\end{picture}\\           
\addtocounter{pics}{-1}\end{center}}
\baselineskip 12pt              

\vbox{
\bpic{400}{120}{20}{-20}

\put(48,20){\framebox(4,3){}}
\put(78,20){\framebox(4,3){}}
\put(54,23){\vector(1,0){22}}
\put(76,20){\vector(-1,0){22}}
\put(47,24){\line(-1,2){50}}
\put(82,24){\line(1,2){50}}


\put(51,51){\circle*{3}}
\put(79,51){\circle*{3}}
\put(54,49){\vector(1,0){22}}
\put(76,52){\vector(-1,0){22}}
\put(49,53){\line(-1,2){30}}
\put(81,53){\line(1,2){30}}

\put(65,-10){\case{$f_{r,s}(\D,\htop,t)=0$, $\Delta \neq - l$}}


\put(198,20){\framebox(4,3){}}
\put(228,20){\framebox(4,3){}}
\put(204,23){\vector(1,0){22}}
\put(226,20){\vector(-1,0){22}}
\put(197,24){\line(-1,2){50}}
\put(232,24){\line(1,2){50}}
\put(201,51){\circle*{3}}
\put(229,51){\circle*{3}}
\put(226,51){\vector(-1,0){22}}
\put(199,53){\line(-1,2){30}}
\put(231,53){\line(1,2){30}}
\put(215,-10){\case{$f_{r,s}(\D, \htop, t)=0$, $\Delta = - l$, 
$t \neq - {s \over n}$}}

\put(348,20){\framebox(4,3){}}
\put(378,20){\framebox(4,3){}}
\put(354,23){\vector(1,0){22}}
\put(376,20){\vector(-1,0){22}}
\put(347,24){\line(-1,2){50}}
\put(382,24){\line(1,2){50}}
\put(351,51){\circle*{3}}
\put(379,51){\circle*{3}}
\put(349,53){\line(-1,2){30}}
\put(381,53){\line(1,2){30}}
\put(353,53){\line(1,2){30}}
\put(377,53){\line(-1,2){30}}
\put(365,-10){\case{$f_{r,s}(\D, \htop, t)=0$, $\Delta = - l$, 
$t = - {s \over n}$ }}
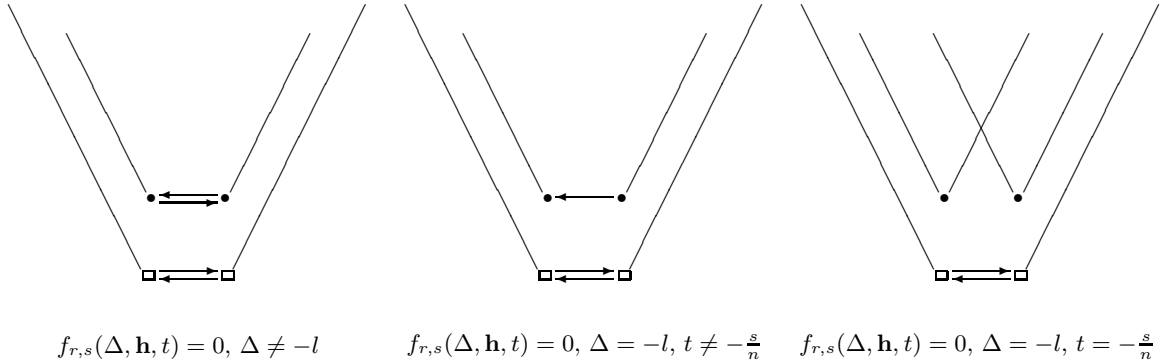
\epic{. The \svec s corresponding to the series $f_{r,s}(\D, \htop, t)=0$
belong to two different types of submodules of the same size but different
shape at the bottom. In the first type, as shown in the figures on the left and 
in the center, the two \svec s at the bottom of the submodules are connected 
by one or two arrows: the action of \Gn\ and/or \Qn , depending on whether
$\D \neq -l$ or $\D = -l, \, t \neq - {s \over n}, \, n=1,..,r $ (for which 
one of the \svec s is chiral). In the second type, corresponding to 
$\D = -l, \, t = - {s \over n}, \, n=1,..,r $, the two \svec s are chiral and 
therefore disconnected from each other, as shown in the figure on the 
right (the overlap of the two subsubmodules is another subsubmodule itself).}}

\baselineskip 16pt
\vskip .2in

The third type of submodules, shown in Fig. II, on the left and in the 
center, corresponds to the charged roots of the vanishing planes 
$g_{k}^{\pm}(\D, \htop, t)=0$. As already pointed out, these submodules 
are smaller than the generic ones, with partition functions given by
$P_k(l-k)$. In the most general case ($\D \neq 0$) the
two \svec s at the bottom of the submodule can be described as 
{\it charged}: positively charged $\kc_l^{(1)G}$ in the $\cG$-sector and  
$\kc_l^{(1)Q}$ in the $\cQ$-sector for $g_{k}^{+}(\D, \htop, t)=0$, and
negatively charged $\kc_l^{(-1)G}$ in the $\cG$-sector and  
$\kc_l^{(-1)Q}$ in the $\cQ$-sector for $g_{k}^{-}(\D, \htop, t)=0$. In 
each case one of the \svec s becomes chiral for $\D=-l$ whereas the other
\svec\ does not. The bottom of these submodules is always connected 
therefore as there is no analog to the `chiral-chiral' case of the 
uncharged \svec s. When the bottom of the submodule is at level zero in
the Verma module, then there is only one \svec , which consequently is chiral. 
These submodules seem to fit well the description of `topological' 
submodules given by Semikhatov and Tipunin. Namely, they correspond to 
the charged roots $g_{k}^{\pm}(\D, \htop, t)=0$, they are smaller than the 
generic `massive' submodules and they have one or two states at the bottom. In
the first case the unique state is charged and chiral (called `topological') while
in the second case the two states are connected through \Gn\ and/or \Qn , 
one of these states being the \Gn-closed charged \svec\ $\kc_l^{(1)G}$ (for 
$g_{k}^+(\D, \htop, t)=0$) or $\kc_l^{(-1)G}$ (for $g_{k}^-(\D, \htop, t)=0$).

Finally, the fourth type of submodules, shown in Fig. II on the right, 
corresponds to the `no-label' submodules built on indecomposable \svec s.
These are the widest submodules, with four \svec s at the bottom.
The indecomposable \svec s are {\it primitive} \svec s that only exist 
for discrete values of $\D, \htop, t$, in generic Verma modules in which
there are intersections, at the same level $l$, of \svec s corresponding 
to the series $f_{r,s}(\D, \htop, t)=0$ with \svec s corresponding to one 
of the series $g_k^{\pm}(\D, \htop, t)=0$, with ${rs \over 2} = k = l$ and 
$\D = -l$. The values of $\ctop$ for which indecomposable  \svec s exist are
$\ctop = {3r-6 \over r}\,$, corresponding to $t={2 \over r}$. These results 
were proved in ref. \cite{DB4} although the existence of indecomposable 
 \svec s was established earlier, in January 1997 in ref. \cite{BJI6}, as 
they were explicitly written down at level 1 (moreover, shortly afterwards 
indecomposable \svec s were written down also at level 2 in ref. \cite{DB1}).

\vskip .5in

\baselineskip 12pt
\vbox{
\bpic{400}{120}{20}{-20}


\put(48,20){\framebox(4,3){}}
\put(78,20){\framebox(4,3){}}
\put(54,23){\vector(1,0){22}}
\put(76,20){\vector(-1,0){22}}
\put(47,24){\line(-1,2){50}}
\put(82,24){\line(1,2){50}}

\put(81,81){\circle*{3}}
\put(99,81){\circle*{3}}
\put(84,79){\vector(1,0){12}}
\put(96,82){\vector(-1,0){12}}
\put(79,83){\line(-1,2){20}}
\put(101,83){\line(1,2){20}}

\put(65,-10){\case{$g^{\pm}_k(\Delta,\htop,t)=0$, $\D \neq -l$}}
\put(198,20){\framebox(4,3){}}
\put(228,20){\framebox(4,3){}}
\put(204,23){\vector(1,0){22}}
\put(226,20){\vector(-1,0){22}}
\put(197,24){\line(-1,2){50}}
\put(232,24){\line(1,2){50}}

\put(231,81){\circle*{3}}
\put(249,81){\circle*{3}}
\put(234,79){\vector(1,0){12}}
\put(229,83){\line(-1,2){20}}
\put(251,83){\line(1,2){20}}

\put(215,-10){\case{$g^{\pm}_k(\Delta,\htop,t)=0$, $\D = -l$}}

\put(348,20){\framebox(4,3){}}
\put(378,20){\framebox(4,3){}}
\put(354,23){\vector(1,0){22}}
\put(376,20){\vector(-1,0){22}}
\put(347,24){\line(-1,2){50}}
\put(382,24){\line(1,2){50}}

\put(361,80){\circle*{3}}
\put(380,78){\circle*{3}}
\put(380,82){\circle*{3}}
\put(399,80){\circle*{3}}
\put(384,78){\vector(1,0){12}}
\put(396,82){\vector(-1,0){12}}
\put(364,82){\vector(1,0){12}}
\put(376,78){\vector(-1,0){12}}
\put(359,83){\line(-1,2){20}}
\put(401,83){\line(1,2){20}}

\put(365,-10){\case{no-label submodules}}
\epic{. 
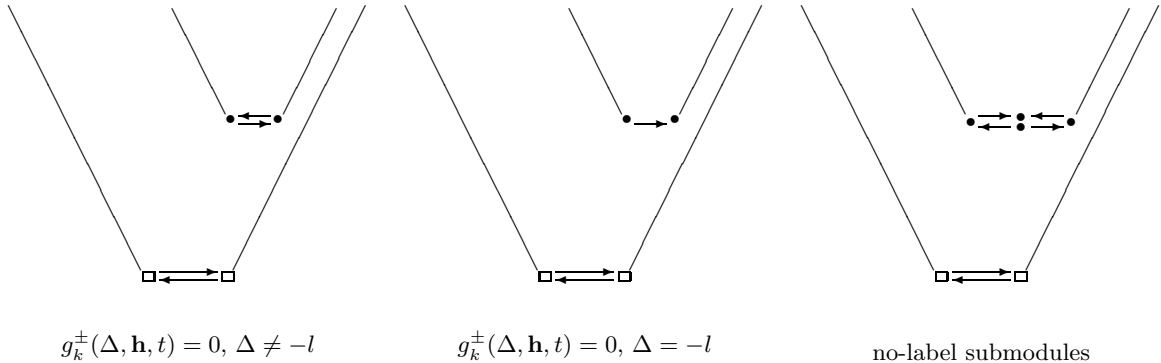
The \svec s corresponding to the series $g_{k}^{\pm}(\D, \htop, t)=0$ 
belong to only one type of submodules, which are smaller than the 
generic ones. In the general case $k \neq 0$ there are two \svec s at the 
bottom of the submodules, connected by \Gn\ and/or \Qn , depending on whether 
$\D \neq -l$ or $\D=-l$. (However, for $k=0$, that is level zero, the bottom 
of the corresponding submodule consists of only one \svec\ which is chiral). 
On the right, the indecomposable \svec s generate the `no-label' submodules,
which are the widest submodules with four \svec s at the bottom. }}

\vskip .2in

\baselineskip 16pt

The action of \Gn\ and \Qn\ on an indecomposable  \svec\ $\kc_l^{(q)}$ produce three 
secondary singular vectors (one \Gn-closed, one \Qn-closed and one chiral)
which cannot `come back' to the no-label \svec\ 
by acting with \Gn\ and \Qn :
 \begin{eqnarray} \Qz \,\kcc{l}{q}{ } \to \kcc{l}{q-1}{Q} ,  
 \quad\Gz \,\kcc{l}{q}{ } \to \kcc{l}{q+1}{G} , \quad 
 \Gz \, \Qz \,\kcc{l}{q}{ } \to \kcc{l}{q}{G,Q} \,. \label {QGnh} 
 \end{eqnarray}
\noi
It happens that one of these secondary \svec s corresponds to 
the series $f_{r,s}(\D, \htop, t)=0$, another one corresponds to the 
series $g_k^{\pm}(\D, \htop, t)=0$, and the remaining one corresponds 
to both series. Hence the bottom of the no-label submodules is connected,
generated by the indecomposable \svec\ and consists of four \svec s: the 
primitive indecomposable  \svec\ and the three secondary \svec s. Obviously, 
these submodules are wider than the `massive' submodules (twice wider 
at the bottom, in fact) and do not fit into the description of `massive' and 
`topological' submodules. The no-label submodules cannot have a counterpart 
in the affine $\hat sl(2)$ algebra simply because there is no $\hat sl(2)$ 
counterpart of the indecomposable N=2 \svec s.

In  Fig. III one can see the case of an uncharged indecomposable \svec\ 
$\ket{\chi}_{l}^{(0)}$ at level l, built on a \Gn-closed h.w. vector 
$\ket{\D, \htop}^G$, with the three secondary \svec s that it generates
by the action of \Gn\ and \Qn .

\vskip .7in
\baselineskip 12pt
\vbox{
\bpic{100}{120}{20}{-20}
\put(48,20){\framebox(4,3){}}
\put(78,20){\framebox(4,3){}}
\put(54,24){\vector(1,0){22}}
\put(76,19){\vector(-1,0){22}}
\put(47,24){\line(-1,2){50}}
\put(82,24){\line(1,2){50}}
\put(65,14){\makebox(0,0){\scriptsize $\cQ_0$}}
\put(65,28){\makebox(0,0){\scriptsize $\cG_0$}}
\put(47,15){\makebox(0,0)[r]{\scriptsize $\ket{-l,\htop-1}^Q$}}
\put(83,15){\makebox(0,0)[l]{\scriptsize $\ket{-l,\htop}^G$}}

\multiput(80,24) (0,5) {14}{\circle*{.5}}
\multiput(78,24) (-2,5) {14}{\circle*{.5}}
\multiput(81,24) (2,5) {14}{\circle*{.5}}

\put(51,91){\circle*{3}}
\put(80,89){\circle*{3}}
\put(80,93){\circle*{3}}
\put(108,91){\circle*{3}}
\put(76,89){\vector(-1,0){22}}
\put(54,93){\vector(1,0){22}}
\put(84,89){\vector(1,0){22}}
\put(106,93){\vector(-1,0){22}}
\put(101,76){\makebox(0,0)[rb]{\scriptsize $\ket{\chi}_{l}^{(0)}$}}
\put(50,89){\makebox(0,0)[rb]{\scriptsize 
	$\ket{\chi}_{l}^{(-1)Q}$}}
\put(116,82){\makebox(0,0)[lb]{\scriptsize 
	$\ket{\chi}_{l}^{(1)G}$}}
\put(60,94){\makebox(0,0)[lb]{\scriptsize 
	$\ket{\chi}_{l}^{(0)G,Q}$}}
\epic{. 
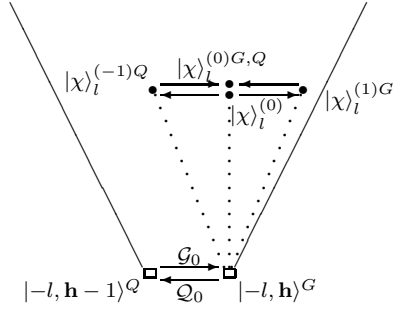
The uncharged indecomposable \svec\ $\ket{\chi}_{l}^{(0)}$ at level $l$, built
on the h.w. vector $\ket{-l,\htop}^G$, is the primitive \svec\ generating
the three secondary \svec s at level $l$:
$\ket{\chi}_{l}^{(1)G} = \cG_0 \ket{\chi}_{l}^{(0)}$,
$\ket{\chi}_{l}^{(-1)Q} = \cQ_0 \ket{\chi}_{l}^{(0)}$ and  
$\ket{\chi}_{l}^{(0)G,Q} = \cQ_0 \cG_0 \ket{\chi}_{l}^{(0)}
= - \cG_0 \cQ_0 \ket{\chi}_{l}^{(0)}$. These cannot generate the indecomposable 
\svec\ by acting with the algebra. However, they are the \svec s detected
by the determinant formula, corresponding to the series 
$f_{r,s}(\D, \htop, t)=0$ ($\ket{\chi}_{l}^{(-1)Q}$ and
$\ket{\chi}_{l}^{(0)G,Q}$) and the series $g_{k}^+(\D, \htop, t)=0$
($\ket{\chi}_{l}^{(1)G}$ and $\ket{\chi}_{l}^{(0)G,Q}$). }}

\baselineskip 16pt
\vskip .4in

The corresponding uncharged indecomposable 
\svec\ $\kc_1^{(0)}$ at level 1, together with the three secondary \svec s 
at level 1 read:
\BE \kc_{1,\ket{-1,-1, \, t=2}^G}^{(0)}\, = 
(\cL_{-1} - \cH_{-1})\ket{-1,-1, \, t=2}^G ,\EE
\BE \kc_{1,\ket{-1,-1, \, t=2}^G}^{(1)G}=
\cG_0\kc_{1,\ket{-1,-1, \, t=2}^G}^{(0)}\, =
 2 \cG_{-1} \ket{-1,-1, \, t=2}^G ,\EE
\BE \kc_{1,\ket{-1,-1, \, t=2}^G}^{(-1)Q}=
\cQ_0\kc_{1,\ket{-1,-1, \, t=2}^G}^{(0)}\,
 = (\cL_{-1}\cQ_0 - \cH_{-1}\cQ_0 - \cQ_{-1})\ket{-1,-1, \, t=2}^G ,\EE 
\BE  
\kc_{1,\ket{-1,-1, \, t=2}^G}^{(0)G,Q} = 
 \cG_0\cQ_0\kc_{1,\ket{-1,-1, \, t=2}^G}^{(0)} = 
 2 (-2\cL_{-1} + \cG_{-1}\cQ_0 )\ket{-1,-1, \, t=2}^G .\EE
\noi
The indecomposable \svec\ only exists for $t=2$ ($\ctop=-3$) whereas 
the three secondary \svec s are just the particular cases, for $t=2$, 
of the one-parameter families of \svec s of the corresponding types, 
which exist for all values of $t$ and were written down in ref. \cite{BJI6}. 

We have shown that the two types of submodules of the Topological N=2  
algebra proposed by  Semikhatov and Tipunin in several papers
-- `massive' and `topological' submodules --  as deduced from the `would 
be isomorphism' between this algebra and the affine $\hat sl(2)$ 
algebra, fit into the `external' description of the submodules of the first 
and third types that we have analyzed. That is, they fit into the description 
as regards size and shape at the bottom/top of the submodule. However,
as we will see in next subsection using an explicit example, these 
submodules do not satisfy a crucial property derived from the `isomorphism': 
they are not generated maximally by the `non-conventional \svec s' 
constructed by the authors. That is, one can find subsingular vectors 
outside the submodules generated by the `non-conventional \svec s'.

For the readers not familiar with the concept of {\it subsingular} vector the
following description can be quite clarifying: A given submodule may not be 
completely generated by the \svec s at the bottom, that is, by the h.w. null vectors. 
These could generate only a subsubmodule of the whole (maximal) submodule, 
in which case one or more subsingular vectors generate the missing parts. 
Whereas the subsingular vectors can reach the \svec s at the bottom by
the action of the generators of the algebra, the contrary is not true: 
subsingular vectors cannot be reached by the action of the algebra on
the  \svec s, therefore they are outside the submodules built on the \svec s.
As a result, when the submodules built on the \svec s are set to zero 
the subsingular vectors surface as new \svec s.

The submodules of the second and fourth types (`chiral-chiral' and 
`no-label' submodules) should not exist were the `isomorphism'
correct. As a matter of fact, there is no $\hat sl(2)$ counterpart for the
indecomposable \svec s that generate the no-label submodules, and
it is not even clear whether there is a $\hat sl(2)$ counterpart for the 
chiral {\it uncharged} \svec s $\ket{\chi}_{l}^{(0)G,Q}$, as they have 
been systematically ignored by the authors. 

Another important remark concerns the presentation of the \svec s of the
Topological N=2 algebra made by the authors, for convenience, in order
to endorse the `isomorphism'. They claim that in the {\it conventional} 
approach the h.w. conditions imposed on the h.w. vectors and on {\it any} \svec\ 
must include the annihilation by \Gn\ (eq.(2.11) in ref. \cite{SeTi3}). This 
statement is not only incorrect but also very misleading. First of all, in the 
conventional approach for the conformal and superconformal algebras,
one defines the h.w. vectors and \svec s (sometimes called simply null vectors) 
as the states with lowest conformal weight (lowest energy) in the Verma 
modules and submodules, respectively. As a result, in most Verma modules and 
submodules of the Ramond and the Topological N=2 algebras (they are 
isomorphic  \cite{DB4}) there are two sectors degenerated in energy, the 
$+$ and $-$ sectors for the Ramond algebra and the $\cG$ and $\cQ$ sectors 
for the Topological algebra, the corresponding states annihilated by the 
fermionic zero modes $G^+_0$ or $G^-_0$ and \Gn\ or \Qn , respectively
\cite{BFK}\cite{KaMa3}\cite{Yu}\cite{Kir1}\cite{LVW}\cite{MSS}\cite{DB4}.  
That is, at the bottom of most Verma modules and submodules of 
the Ramond and of the Topological N=2 algebras there are two h.w. vectors and 
two \svec s, respectively, the fermionic zero modes interpolating between them.
In addition, one can find indecomposable \svec s not annihilated by any of the 
fermionic zero modes, that also must be called \svec s following the conventional 
definition and generate the widest submodules, as we have just shown
\cite{BJI6}\cite{DB1}\cite{DB2}\cite{DB4}.

Second, and this is a crucial point, to break the symmetry between the $\cG$ 
and the $\cQ$ sectors, regarding the \svec s of the $\cQ$-sector simply as 
descendant states of `the \svec s' of the $\cG$-sector, leads to a great deal of 
confusion in the case of zero conformal weight $\D+l=0$. The reason is that 
for $\D+l=0$ the \Qn-closed \svec s $\ket{\chi}_{l=-\D}^{(q)Q}$ are in fact the 
primitive \svec s generating the {\it secondary} \svec s of the $\cG$-sector, 
which are necessarily chiral of type $\ket{\chi}_{l=-\D}^{(q+1)G,Q}$ (see the 
details in ref. \cite{DB4}, Appendix A). In the conventions used by 
Semikhatov and Tipunin, however, the vectors $\ket{\chi}_{l=-\D}^{(q)Q}$ are 
not singular by definition, although they are necessarily null. As a result, since
they are not descendant states of `the \svec ' $\ket{\chi}_{l=-\D}^{(q+1)G,Q}$, 
but the other way around, the \svec s of the $\cQ$-sector $\ket{\chi}_{l=-\D}^{(q)Q}$ 
must be called {\it subsingular vectors} instead. For similar reasons,
the indecomposable  \svec s must also be called subsingular vectors as 
they are null, not descendants of the \svec s of the $\cG$-sector, but
the other way around, and they are not singular by definition.

\vskip 0.5in

\subsection{The classification of subsingular vectors}\lvm

Now we will see that the explicit examples of subsingular vectors of the
Topological N=2 superconformal algebra given
in ref. \cite{BJI6}, which are singular in the chiral Verma modules, do not 
fit into the `complete classification' of subsingular vectors presented
in ref. \cite{SeTi3}. As a bonus we will also deduce 
that the non-conventional \svec s constructed in refs. \cite{SeTi1} \cite{SeTi2}
do not generate maximal submodules, contrary to the claims of the authors
who deduced this property directly from the `isomorphism'. This property,
in addition, was used as a major tool for the classification of the subsingular 
vectors. 

The authors classified the generic Verma modules built on \Gn-closed
h.w. vectors (`massive' Verma modules) 
according whether they have zero, one, two or more \svec s from the 
uncharged and/or charged series associated to the roots of the determinant 
formula (in our notation $f_{r,s}(\D, \htop, t)=0$ and/or 
$g_k^{\pm}(\D, \htop, t)=0$, eqns. (2.2) and (2.3)). 
In every case they applied the assumption 
that there are only two types of submodules -- `massive' and `topological' 
-- and these are generated maximally by the non-conventional \svec s
constructed in refs. \cite{SeTi1}\cite{SeTi2}. 
Namely, one `twisted topological' non-conventional \svec\
(where they mean twisted by the spectral flows\footnote{In our opinion,
the authors used the spectral flows in a very dubious way. Apart from
the `classical' reference on the subject \cite{SS}, the interested reader
may also find useful the analysis done in refs. \cite{BJI3} and \cite{B1}.}) 
is assumed to generate maximally one `topological' submodule 
whereas one `twisted massive' non-conventional \svec\ 
is assumed to generate maximally one `massive' submodule.\footnote{
As was mentioned before, if these non-conventional \svec s could
generate maximally, by the action of the algebra, the whole submodules
whereas the conventional h.w. singular vectors failed to do the same,
then the non-conventional \svec s would be nothing but subsingular vectors
or some of their descendants, equivalently.}

These non-conventional \svec s are null states that in general are 
not located at the bottom of the submodules unlike the conventional
 \svec s. In fact, in the cases when
they lie at the bottom then they coincide with the conventional \svec s.   
An important remark is that the `twisted topological' h.w. conditions satisfied
by the `twisted topological' non-conventional \svec s reduce to the
chirality h.w. conditions (i.e. annihilation by \Gn , \Qn\ and by all the
positive generators) in the case of the twist parameter equal to zero.
As a result, the 'zero twist topological' non-conventional \svec s
coincide with the chiral {\it charged} conventional \svec s at the
bottom of the `topological' submodules. 

Using these assumptions and simple geometrical arguments the authors
deduced in which cases the {\it conventional} \svec s at the bottom of the
submodules do not generate maximal submodules, and then using the
non-conventional \svec s they `identified' the subsingular vectors, giving
some general expressions in some cases.
The subsingular vectors given by us in ref. \cite{BJI6} corresponded
necessarily to the ones described by the authors in the case
`codimension-2 charge-massive', given\footnote{The authors themselves
claimed that the subsingular vectors given in ref. \cite{BJI6} were
described by Proposition 3.9, case $n=0$, although they did not 
explicitely mention this in the last revised, published version of
ref. \cite{SeTi3}.}
in Proposition 3.9, for $n=0$, since they are located in Verma modules
with one charged chiral \svec\ (at level zero, what gives $n=0$) and one 
uncharged \Gn-closed \svec . In the notation of the authors, who draw 
the Verma modules from the top downwards,
the charged \svec\ is both a conventional `top-level' \svec\ and a 
non-conventional `twisted topological' charged \svec\ 
$\ket{E(n)}_{ch}$ with twist parameter $n=0$ 
(i.e. the non-conventional \svec\ is at the bottom of the submodule,
in our notation, so that it coincides with the conventional \svec ). 
The uncharged \Gn-closed \svec\ is described as the conventional 
`top-level' uncharged \svec\ in the `massive' submodule generated 
by the non-conventional `massive' \svec\  $\ket{S(r,s)}$, and is 
denoted as $\ket{s}$.

For this case, and in fact for all cases `described' by Proposition 3.9,
the authors deduced that a subsingular vector $\ket{Sub}$ must exist
{\it inside} the maximal massive submodule generated by $\ket{S(r,s)}$
in the sense that $\ket{Sub}$ is located {\it outside} the non-maximal
submodule generated by the conventional uncharged \svec\ $\ket{s}$, 
becoming singular once $\ket{s}$ is set to zero. This implies
that the subsingular vector $\ket{Sub}$ is `pushed down' (`up' in the 
authors figures) by the action of the lowering operators inside the 
non-maximal submodule generated by $\ket{s}$, so that setting
this submodule to zero is equivalent to push down the vector to nothing,
i.e. the subsingular vector becomes singular. Observe that in this case
the subsingular vector $\ket{Sub}$, once it reaches $\ket{s}$ by
the action of the lowering operators, cannot go down (`up') 
anymore since $\ket{s}$ is the conventional \svec\ at the bottom of
the submodule annihilated by all the lowering operators. In other words,
if the subsingular vector $\ket{Sub}$ becomes singular when $\ket{s}$ is 
set to zero, then acting with the lowering operators on $\ket{Sub}$ it
cannot be pulled down beyond the level of $\ket{s}$, getting in fact 
`stuck' in $\ket{s}$ (up to constants).

The subsingular vectors at level 3 given by us in ref. \cite{BJI6} do not
follow the behaviour described by Proposition 3.9, however. Rather, they
are pulled down beyond the uncharged conventional \svec\ $\ket{s}$ 
that one finds at level 2 and, in fact, they can be pulled down until
the very end, i.e. level zero, becoming singular only when the charged 
chiral \svec\ $\ket{E(0)}_{ch}$ at level zero is set to zero. As a
consequence, these subsingular vectors do not become singular when 
$\ket{s}$ is set to zero, what implies that they are not pulled inside the 
submodule generated by $\ket{s}$ by the action of the lowering operators
(see Fig. IV), 
and therefore they are not located inside the maximal massive submodule 
supposed to be generated by the `massive' \svec\ $\ket{S(r,s)}$. But these 
subsingular vectors are neither located inside the submodule generated 
by $\ket{E(0)}_{ch}$ since they do not disappear when $\ket{E(0)}_{ch}$
is set to zero, becoming singular rather. In other words, 
as shown in Fig. IV, these subsingular
vectors are pulled inside the submodule generated by $\ket{E(0)}_{ch}$
by acting with the lowering operators. This implies that the submodule
generated by the non-conventional \svec\ $\ket{E(0)}_{ch}$ is not maximal,
in contradiction with the claims in ref. \cite{SeTi3}.   
 
\vskip .8in
\baselineskip 12pt
\vbox{
\bpic{250}{120}{20}{-20}



\put(100,20){\circle*{.5}}
\put(128,20){\framebox(4,3){}}
\put(126,20){\vector(-1,0){22}}
\put(97,24){\line(-1,2){50}}
\put(132,24){\line(1,2){50}}
\put(115,14){\makebox(0,0){\scriptsize $\cQ_0$}}
\put(97,15){\makebox(0,0)[r]{\scriptsize $\ket{E(0)}_{ch} = \cQ_0\ket{0,2}^{G}$}}
\put(133,15){\makebox(0,0)[l]{\scriptsize $\ket{0,2}^{G}$}}
\put(101,24){\line(1,6){16}}
\put(99,24){\line(-1,3){32}}

\put(103,80){\makebox(0,0)[rb]{\scriptsize $\ket{s}$}}
\multiput(100,75) (1,1) {39}{\circle*{.5}}
\multiput(98,73) (-1,0) {9}{\circle*{.5}}
\multiput(88,75) (-1,2) {18}{\circle*{.5}}
\put(138,113){\circle*{3}}
\put(155,115){\makebox(0,0)[rb]{\scriptsize $\ket{S(r,s)}$}}


\put(129,87){\circle*{3}}
\put(126,85){\vector(-1,-1){29}}
\put(159,89){\makebox(0,0)[rb]{\scriptsize $\ket{Sub}_3^{(1)}$}}
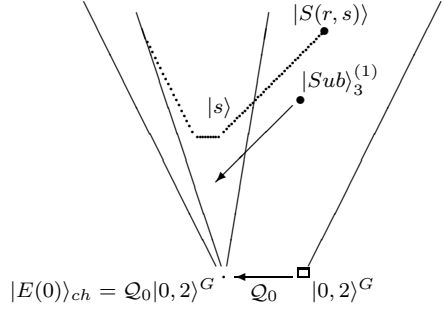
\epic{. When the charged level zero \svec\ $\ket{E(0)}_{ch} = \cQ_0\ket{0,2}^G$
is set to zero, the generic (`massive') Verma module built on $\ket{0,2}^G$ is divided 
by the submodule generated by this \svec . As a result one obtains the chiral 
(`topological') Verma module built on the chiral h.w. vector $\ket{0,2}^{G,Q}$. 
The subsingular vector $\ket{Sub}_3^{(1)}$ at level 3 is 
outside the submodule generated by $\ket{E(0)}_{ch}$, being
pulled inside by the action of the lowering operators. 
Consequently, the submodule generated by the non-conventional `topological' 
charged \svec\ $\ket{E(0)}_{ch}$ (which being at the bottom of the
submodule coincides with the conventional chiral \svec\ $\cQ_0\ket{0,2}^G$)
is not maximal since there is (at least) one subsingular vector left outside.
This subsingular vector becomes singular, therefore,
in the chiral Verma module $V(\ket{0,2}^{G,Q})$ obtained after the quotient.
Inside the submodule generated by $\ket{E(0)}_{ch}$ one finds the 
uncharged \Gn-closed \svec\ $\ket{s}$ (and its companion in the 
$\cQ$-sector that is not indicated). The subsingular vector $\ket{Sub}_3^{(1)}$
is not pulled inside the submodule generated by  $\ket{s}$ by the lowering 
operators and therefore it does not become singular once $\ket{s}$ is set to zero. 
Rather, it is pulled down to lower levels than $\ket{s}$. As a result 
$\ket{Sub}_3^{(1)}$ does not belong to the `massive' submodule, 
supposedly to be maximally generated by the non-conventional
`massive'  \svec\ $\ket{S(r,s)}$, having $\ket{s}$ at the bottom. }}

\vskip .18in
\baselineskip 16pt

One example given in ref. \cite{BJI6} is the subsingular vector 
$\ket{Sub}_3^{(1)}$ at level 3 with charge $q=1$ built on the \Gn-closed
h.w. vector $\ket{\D, \htop}^G$ with conformal weight $\D=0$ and U(1) 
charge $\htop=2$:
\begin{eqnarray*}
\ket{Sub}_3^{(1)}= \{ {3-\ctop\over 24} {\ } \cL_{-1}^2 \cG_{-1} - {3\over 4} {\ }
\cL_{-1} \cG_{-2} - {1\over 4} {\ } \cL_{-2} \cG_{-1} + 
{\ctop+9\over 4(\ctop-3)} {\ } \cH_{-2} \cG_{-1} +  \\
{27 - \ctop\over 4(3-\ctop)}  {\ } \cG_{-3} + {6\over\ctop-3} {\ } \cH_{-1} \cG_{-2} + 
{3\over 4} {\ } \cH_{-1} \cL_{-1} \cG_{-1} + 
{3\over 3-\ctop} {\ } \cH_{-1}^2 \cG_{-1} \} {\ } \ket{0,2}^G \,.
\end{eqnarray*}
\noi
Acting with $\cQ_1$ on this vector one does not hit the conventional
uncharged \svec\ $\ket{s}$ at level 2 but one reaches the state
\BE  \{ {\ctop-12\over 12} {\ } \cL_{-1} \cG_{-1} + 
{3(11-\ctop)\over 4(3-\ctop)}  {\ } \cG_{-2} +
{3(11-\ctop)\over 4(\ctop-3)}  {\ } \cH_{-1} \cG_{-1} \} {\ } \cQ_0 {\ } \ket{0,2}^G \,,
\EE
\noi
which is a non-singular descendant of the level zero charged \svec\  
$\,\ket{E(0)}_{ch} = \cQ_0 \, \ket{0,2}^G $. That is, $\ket{Sub}_3^{(1)}$ is
pulled inside the submodule generated by $\ket{E(0)}_{ch}$ by the 
action of $\cQ_1$. Acting further with $\cL_1$ 
one reaches the state $\, \cG_{-1} \cQ_0 \, \ket{0,2}^G \,$ at level 1
which, again, is not singular. Acting with $\cQ_1$ on this state one
reaches finally the level zero chiral charged \svec :
$\,\cQ_1\cL_1\cQ_1 \, \ket{Sub}_3^{(1)} = \cQ_0 \, \ket{0,2}^G = \ket{E(0)}_{ch}$.

This example not only proves that Proposition 3.9 is incorrect, as 
$\ket{Sub}_3^{(1)}$ does not become singular when $\ket{s}$ is set to  
zero, and that the subsingular vectors presented in ref. \cite{BJI6}
(the only examples known at that time!) do not fit into the
`complete' classification of subsingular vectors given 
in ref. \cite{SeTi3}. As we have just discussed, this
example also proves that the non-conventional topological \svec\ 
$\,\ket{E(0)}_{ch} = \cQ_0  \ket{0,2}^G $ (which is located at the
bottom of the submodule and therefore coincides with the conventional 
chiral \svec ) does not generate a maximal submodule since the
subsingular vector $\ket{Sub}_3^{(1)}$ is outside this submodule, 
being pulled inside by the action of the lowering operators. 
This example disproves the claim  that
the non-conventional `massive' and `topological' \svec s generate
maximal submodules, i.e;  with no space left outside for subsingular vectors.
Indeed, we have shown that the subsingular vector           
$\ket{Sub}_3^{(1)}$ is neither generated by the `massive' \svec\ 
$\ket{S(r,s)}$ nor by the `topological' \svec\ $\ket{E(0)}_{ch}$, nor
by both of them together.

\vskip .5in

{\bf Acknowledgements}\lvm

\small

We thank A. Belavin for discussions. This work has been partially 
supported by the Spanish Ministerio de Ciencia e Innovaci\'on,  
Research Project FPA2005-05046, and by the Project 
CONSOLIDER - INGENIO 2010, Programme CPAN (CSD2007 - 00042).

\end{document}